\documentclass{article}
\usepackage[utf8]{inputenc}
\usepackage[margin=1in]{geometry}
\usepackage[titletoc,title]{appendix}
\usepackage{hyperref}
\usepackage{import}
\usepackage{amsmath,amsfonts,amssymb,mathtools,amsthm}
\usepackage{animate}
\usepackage{multicol}
\usepackage{lipsum}
\usepackage{graphicx}
\usepackage{subcaption}
\usepackage{mwe}
\usepackage{soul}
\usepackage{xcolor}
\usepackage{comment}
\newcommand{\R}{\mathbb{R}}
\newcommand{\C}{\mathbb{C}}

\newcommand{\Z}{\mathbb{Z}}
\newtheorem{theorem}{Theorem}[section]

\newtheorem{definition}{Definition}[section]
\newtheorem{proposition}{Proposition}
\newtheorem{conjecture}{Conjecture}
\newtheorem{lemma}{Lemma}
\newcommand{\pd}[2]{\frac{\partial #1}{\partial #2}}

\newcommand{\norm}[1]{\left\lVert#1\right\rVert}

\title {The Hitchin and Sutcliffe metrics for hyperbolic 2-monopoles}
\author{Thomas Galvin}
\author{
  Thomas Galvin \\
  School of Mathematics, University of Leeds \\
  \texttt{mmtga@leeds.ac.uk}
}

\begin{document}
\begin{titlepage}
\maketitle
\thispagestyle{empty}
\begin{abstract}
    We compare two currently known examples of hyperbolic SU(2) two-monopole metrics and their candidacy as models of hyperbolic monopole dynamics. The first is Sutcliffe's boundary metric and the other is a well-known self-dual Einstein metric found by Hitchin. We show that these metrics are conformally inequivalent and make significantly different predictions for monopole dynamics. We compare each metric to the Franchetti--Ross point particle approximation of hyperbolic monopole dynamics. We show Sutcliffe's metric does not converge to the point particle approximation in the large separation limit. We also check Sen's conjecture for each metric by computing their $L^2$ harmonic two-forms. By a symmetry argument we are able to show that Hitchin's metric has a single $L^2$ harmonic two-form consistent with the conjecture.
\end{abstract}
\end{titlepage}
\tableofcontents
\newpage
\section{Introduction}
In this article we explore the problem of describing the dynamics of hyperbolic 2-monopoles. This is currently an open problem, solutions of which have been proposed by both Sutcliffe in \cite{sutcliffe2022hyperbolic} and Hitchin in \cite{hitchin1993new}. The methods we employ to understand the dynamics of hyperbolic monopoles are by analogy with the dynamics of Euclidean monopoles, which we briefly explain below. In \cite{manton1982remark} Manton argued that the dynamics of slowly moving monopoles are well approximated by geodesics on the moduli space -- the metric and moduli space are described below. This claim was made rigorous and then proven by Stuart in \cite{stuart1994geodesic}. This is known as the geodesic approximation, and has motivated significant interest in finding metrics on monopole moduli space. The most famous example is due to Atiyah and Hitchin \cite{atiyah2014geometry} where they found the $L^2$ metric for charge $2$ Euclidean monopoles.

A hyperbolic monopole is a pair $(A,\Phi)$ where $\Phi$ is a $\mathfrak{su}(2)$-valued field and $A$ an $\mathfrak{su}(2)$ 1-form each over hyperbolic 3-space. These fields must satisfy the Bogomolny equation
\begin{align}\label{eqn :: bog}
    \star d_A \Phi = F_A 
\end{align}
with boundary condition
\begin{align}\label{eqn :: mass}
    \norm{\Phi} \to m
\end{align}
as the geodesic distance $\rho \rightarrow\infty$ and $m\in \R_{>0}$. Unless otherwise stated, we take $m=1/2$ for the remainder of this article.

There is a topological invariant called the charge denoted by $N$ which we define as the topological degree of the map
\begin{align}
    \Phi|_{\rho = \infty} : \partial H^3 \to S^2 \subset \mathfrak{su}(2)
\end{align}
since $\partial H^3 \cong S^2$.

The moduli space of hyperbolic monopoles, defined as the solutions to \eqref{eqn :: bog} subject to \eqref{eqn :: mass} up to based gauge transformations, forms a smooth manifold of dimension $4N-1$. In the Euclidean case, it is common to take the quotient by the action of translations to define the moduli space of centred monopoles; we denote this by $\mathcal{M}^\mathbb{E}_N$. An analogous notion for hyperbolic monopoles was given by Murray, Norbury, and Singer in \cite{murray2001hyperbolic}, though it is not as straightforward to state. However, in the specific case of charge $N=2$, we can bypass this complication by restricting our focus to the submanifold of inversion-symmetric 2-monopoles, which we denote as $\mathcal{M}_2$.

Now we can define Manton's metric on the moduli space of Euclidean monopoles. The metric is defined as the $L^2$ norm of the zero modes. Zero modes are just solutions to the linearised Bogomolny equations at a point $(A,\Phi)$ on the moduli space of charge $N$ Euclidean monopoles:
\begin{align}
\star d_Aa - d_A\phi + [\Phi,a] = 0
\end{align}
where $a$ is a Lie algebra value one-form and $\phi$ a Lie algebra valued function. The zero modes $(a,\phi)$ must be orthogonal to the gauge orbits which can be achieved by imposing the Coulomb gauge fixing condition: 
\begin{align}\label{eq::gaugefixing}
d^\dagger_Aa + [\Phi,\phi] = 0.
\end{align}
Then the metric on $\mathcal{M}^\mathbb{E}_N$ is given by
\begin{align}\label{Metric}
h := \int_{\R^3}||a||^2+||\phi||^2.
\end{align}
This is known as the $L^2$ metric and is known to be a complete Riemannian metric on the moduli space of Euclidean monopoles \cite{atiyah2014geometry}.

Atiyah and Hitchin exploited the geometry of the moduli space to find the metric on the centred 2-monopole space which is $SO(3)$ invariant with the general form
\begin{align}
h = \Lambda_0(r)^2dr^2 + \Lambda_1(r)^2\sigma_1^2 + \Lambda_2(r)^2\sigma_2^2 + \Lambda_3(r)^2\sigma_3^2
\end{align}
where $\sigma_i$ are the left-invariant one-forms and the parameter $r$ can be roughly interpreted as the separation between each monopole and $\Lambda_i(r)$ are smooth functions.

Unfortunately, for the hyperbolic case the $L^2$ metric of the zero-modes is known to diverge and so this cannot define a metric on the hyperbolic moduli space. One resolution to this problem has been suggested by Sutcliffe in \cite{sutcliffe2022hyperbolic} where he found the divergent hyperbolic version of the integral \eqref{Metric} can be regularized. To find this metric Sutcliffe made use of the fact that hyperbolic monopoles are uniquely determined by their associated abelian connection on the $S^2$ boundary of hyperbolic space. The $L^2$ metric of these fields can be computed and agrees with the regularized $L^2$ metric. The only other explicit metric computed on the moduli space of hyperbolic 2-monopoles, known to the author, is due to Hitchin in \cite{hitchin1993new} though he was not directly interested in hyperbolic monopoles but instead in self-dual Einstein metrics. The first thing we show in section \ref{sec :: conf ineq} is that these metrics make significantly different predictions for monopole dynamics as we show they do not agree asymptotically or conformally. 

Another aspect of Euclidean monopole dynamics is the agreement, asymptotically with respect to separation, between the $L^2$ metric and a point particle approximation due to Gibbons and Manton in \cite{gibbons1995moduli}. The well separated $N$-monopole $L^2$ metric was shown rigorously to agree asymptotically with the Gibbons-Manton metric by Bielawski in \cite{bielawski1998monopoles}. The equivalent point particle approximation was found by Franchetti and Ross in \cite{franchetti2023asymptotic} for hyperbolic monopoles, where they found that for two monopoles the point particle approximation gives the hyperbolic Taub-NUT metric (hTN). In section \ref{sec:: pointp} we compare the two candidate hyperbolic monopole metrics to the point particle metric. Neither metric agrees entirely with the point particle picture: Hitchin's metric agrees asymptotically and up to a conformal rescaling with hTN \cite{franchetti2023asymptotic}, in contrast to Sutcliffe's metric which does not. The fact that Hitchin's metric agrees asymptotically with the point particle approximation after a conformal rescaling suggests a different metric in the conformal class should be chosen to model hyperbolic monopole dynamics. Another reason to choose a conformally equivalent metric to Hitchin's metric is that it is incomplete due to the presence of a conical singularity. However, one would expect a metric describing monopole dynamics to be complete. In section \ref{sec:: dyn} we compare the geodesics of hTN with from those coming from Sutcliffe's metric in the large separation limit. We find asymptotic agreement with respect to the motion of monopole position but strong disagreement with motion in the relative phase. 

A possible resolution not explored in this article is the recent work of Franchetti and Harland in \cite{franchetti20242}. They considered a different gauge fixing condition to that in equation \eqref{eq::gaugefixing} introduced by Figueroa-O’Farrill and Gharamti in \cite{figueroa2014supersymmetry}. This new gauge fixing condition is given by
\begin{align}
    d^\dagger_A a + [\Phi,\phi] + 2is\phi = 0
\end{align}
where $-s^2$ is the sectional curvature. The resulting $L^2$ metric was shown by Franchetti and Harland to be finite and, in the case of inversion symmetric monopoles, real. The $L^2$ metric for 2-monopoles with mass parameter $1/2$ and its asymptotics will be studied elsewhere \cite{franchetti2026}. However, it might finally provide a completely satisfactory hyperbolic analogue of the Atiyah-Hitchin metric.

Another aspect of Euclidean monopoles is their quantization. In \cite{sen1994dyon} A. Sen used a supersymmetric quantization to show that bound states of Euclidean 2-monopoles are given by $L^2$ harmonic 2-forms and showed that the space of such forms is one dimensional. Sen made a conjecture on the dimension of these forms for any charge $N$. This conjecture was formalized mathematically by Segal and Selby in \cite{segal1996cohomology}. This gives us another avenue in which to probe the metrics of Hitchin and Sutcliffe: we can study their quantum mechanics by computing their respective harmonic forms. In section \ref{sec:: harm} we find that for Hitchin's metric, the space of $L^2$ harmonic two-forms is also one-dimensional, consistent with Sen's predictions for the Euclidean case.

The outline of this article is as follows. In section 2 we discuss some general background information regarding the Riemannian metrics on 2-monopole moduli spaces and introduce the metrics of Sutcliffe and Hitchin. In section 3 we consider whether or not these metrics agree at all with each other asymptotically and/or up to some conformal rescaling. In section 4 we review the work of Franchetti and Ross regarding point particle dynamics and see to what extent Hitchin's and Sutcliffe's metrics agree with this approximation. In Section 5 we consider the geodesics in the large monopole separation limit of Sutcliffe's metric and by considering conserved quantities derive the magnetic-geodesic equation for both Sutcliffe's metric and hTN. Finally, in section 6 we derive the harmonic forms for Sutcliffe's and Hitchin's metrics respectively and compare this to the predictions of Sen concerning Euclidean monopoles.
\section{The Metrics}

The metrics we consider are defined on the moduli space of hyperbolic 2-monopoles modulo gauge transformations. By restricting to the space of inversion symmetric monopoles we fix a centre of mass. By a result of Murray and Singer in \cite{murray1996spectral} we can identify this space as the space of inversion symmetric spectral curves of hyperbolic monopoles. By a result of Norbury and Romao \cite{norbury2007spectral} any such curve can be specified by a mass parameter $m\in \R_{>0}$, a real parameter $0\le r < 1$ and a rotation in $SO(3)\subset SL(2,\C)$. When $m=1/2$, then any spectral curve takes the form  
\begin{align}\label{eqn :: the spectral curves}
    -r(\eta^2\zeta^2 + 1) + (\eta^2 + \zeta^2) - (1-r^2)\eta\zeta = 0
\end{align}
and $SO(3)$ orbits; this is equation 3.14 in \cite{sutcliffe2022hyperbolic}.

The stabilizer of \eqref{eqn :: the spectral curves} under the action of $SO(3)$ is $\Z_2\times\Z_2$ corresponding to $\pi$ rotations about each of the three axes of $H^3$. In the $r = 0$, axially symmetric case, the spectral curve becomes singular and the stabilizer increases to $O(2)$ with degenerate orbits $SO(3)/O(2) \cong \R P^2$. We refer to this degenerate orbit as a bolt, in all cases we consider these bolts give rise to removable singularities in the metrics. Then the moduli space of smooth spectral curves can be identified with a closure of
\begin{align}
    \mathcal{M}_2 \cong (0,1)\times SO(3)/\Z_2\times \Z_2.
\end{align}
Any Riemannian metric on this space will lift to a metric on $(0,1)\times SO(3)$. All the metrics we consider are of the following form:
\begin{align}
    g = f(s)^2ds^2 + a(s)^2\sigma_1^2 + b(s)^2\sigma_2^2 + c(s)^2\sigma_3^2 
\end{align}
where $\sigma_i$ are the left invariant one-forms obeying $d\sigma_1 = \sigma_2\wedge\sigma_3$ and cyclic permutations and $s$ takes values in the open interval $(0,1)$. We can orient such a metric so that the bolt occurs when $s=0$ and near this point they take the form:
\begin{align}
    g \sim ds^2 + 4s^2\sigma_1^2 + C(\sigma_2^2 + \sigma_3^2).
\end{align}
Our conventions will be that the coefficient of $\sigma_1^2$ will always have this form. There is a natural action of $U(1)$ acting on the spectral curve as $g_\varphi\cdot (\eta, \zeta) = (e^{i\varphi}\eta, e^{i\varphi}\zeta)$ where $\varphi \in [0,2\pi]$ - corresponding to rotations about the axis containing the two monopole zeros. The generator of this action, $X$ say, will be a left-invariant vector field of $SO(3)$. We further restrict our conventions so that 
\begin{align}
    \sigma_i(X) = \delta_{i,3}.
\end{align}

\subsection{Sutcliffe's Metric}

In \cite{braam1990boundary} Braam and Austin defined a metric on the hyperbolic $N$-monopole moduli space of mass $m\in \frac{1}{2}\Z$ using the abelian connection of the hyperbolic monopole. Their insight was that the fields on the boundary $S^2_\infty$ determine the hyperbolic monopole and then they used this boundary data to define their moduli space and in turn a metric on it. In \cite{sutcliffe2022hyperbolic} Sutcliffe exploited an observation of Murray, Norbury and Singer in \cite{murray2001hyperbolic} which was that the boundary field $A_\infty$ can be computed as a Chern connection on the anti-diagonal $\Bar{\Delta} \cong \C P^1 \subset \C P^1\times\C P^1$. More explicitly, we define a hermitian metric on the line bundle $L^-$ given locally by
\begin{align}\label{Hermy}
h(z,\bar{z}) := \bar{z}^Np(-\bar{z}^{-1},z),
\end{align}
where $p$ is a local section of $\mathcal{O}(2,2)$ defining the spectral curve. The compatible connection is then 
\begin{align}\label{eqn :: boundary gauge field}
A_\infty =  \partial_{z}\log{h}.
\end{align}
Sutcliffe then computes the $L^2$ metric of these fields over the family of curves given by equation \eqref{eqn :: the spectral curves} (equation 3.14 in \cite{sutcliffe2022hyperbolic}). The metric is then the usual $L^2$ metric on the boundary fields. In \cite{sutcliffe2022boundary} Sutcliffe relates the $L^2$ metric on the fields defined by equation \eqref{eqn :: boundary gauge field} to a renormalized integral corresponding to the hyperbolic analogue of the $L^2$ metric on the fields in the bulk, which as we have already mentioned would otherwise diverge. We sketch these ideas below. Let $A,\Phi$ be a charge $N$ hyperbolic monopole with boundary conditions given by
\begin{align}
    \Phi &\sim \begin{pmatrix}
        im & 0 \\
        0   & -im
    \end{pmatrix}\\
    A &\sim \begin{pmatrix}
        A_\infty & 0 \\
        0   & -A_\infty
    \end{pmatrix}
\end{align}
near $\partial H^3 \cong S^2$. Let $a,\phi$ be the corresponding tangent fields satisfying the linearized Bogomolny equation as well as the Coulomb gauge fixing condition \eqref{eq::gaugefixing}.

In the hyperbolic ball model of sectional curvature $-1$ we have
\begin{align}
    ds^2 = \frac{4(dx^2 + dy^2 + dz^2)}{(1-|x|^2)^2}
\end{align}
Then the $L^2$ metric component $g_{L^2}(a,a)$ diverges logarithmically as $|x|\to 1$. Let $R := |x|$, following \cite{sutcliffe2022boundary} we define a regularized integral $T_b$ as follows:
\begin{align}
    g_{L^2}(a,a) &= \lim_{b\to1} T_b, \quad \text{where}\\
    T_b &= \int_{r<b} (|a_x|^2+|a_y|^2+|a_z|^2) \frac{2R^2 \sin{\theta}dRd\theta d\varphi}{1-R^2} \sim \mathcal{O}(\log{(1-b)}) 
\end{align}
We renormalize by taking the limit
\begin{align}
    \hat{T} = \lim_{b\to 1} \frac{T_b}{\log{(1-b)}} = \int_{S^2} (|a_x|^2+|a_y|^2+|a_z|^2) \sin \theta d\theta d\varphi.
\end{align}
Furthermore, on the $S^2$ boundary the gauge fixing condition \eqref{eq::gaugefixing} reduces to
\begin{align}
    d_{A} \star a = 0
\end{align}
which implies $a_R = 0$ and the remaining field components are divergence free. This is just the abelian Coulomb gauge condition on $S^2$ and by the result of Braam and Austin these fields determine the monopole uniquely.

\begin{figure}[ht]
        \centering
        \begin{subfigure}[b]{0.475\textwidth}
            \centering
            \includegraphics[width=\textwidth]{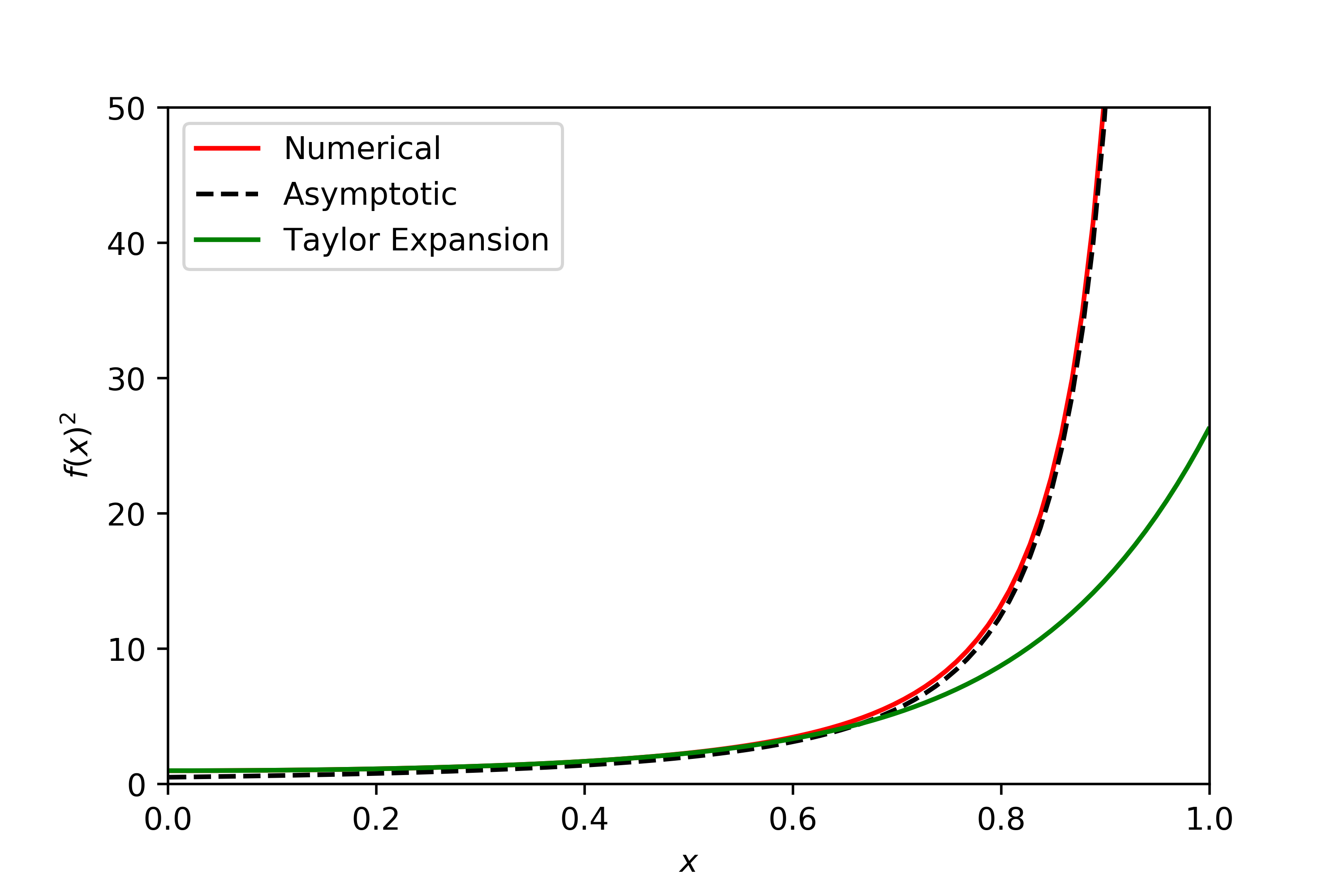}
            \caption[fsq]%
            {{\small $f(x)^2$}}    
            \label{fig:mean and std of net14}
        \end{subfigure}
        \hfill
        \begin{subfigure}[b]{0.475\textwidth}  
            \centering 
            \includegraphics[width=\textwidth]{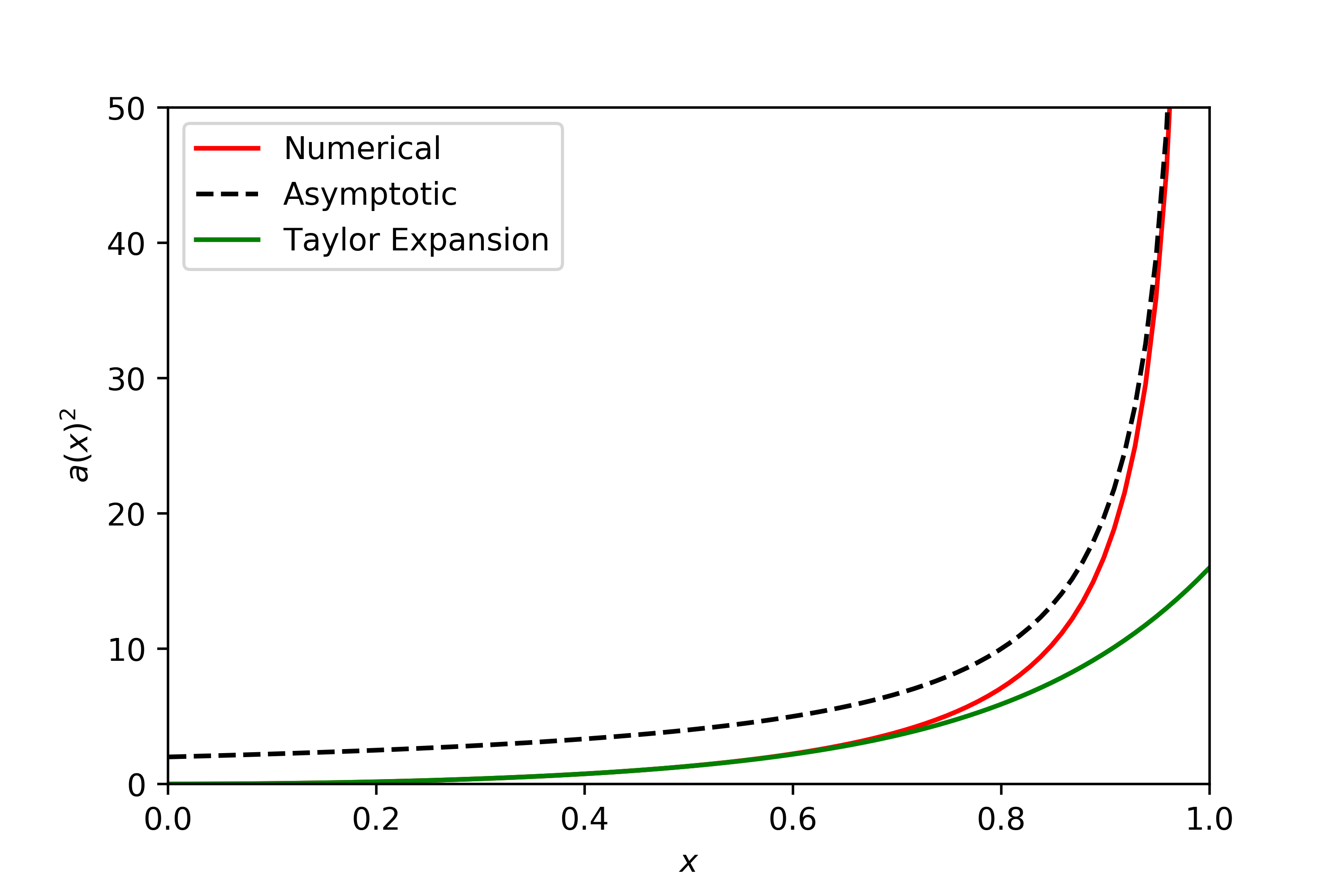}
            \caption[asq]%
            {{\small $a(x)^2$}}    
            \label{fig:mean and std of net24}
        \end{subfigure}
        \vskip\baselineskip
        \begin{subfigure}[b]{0.475\textwidth}   
            \centering 
            \includegraphics[width=\textwidth]{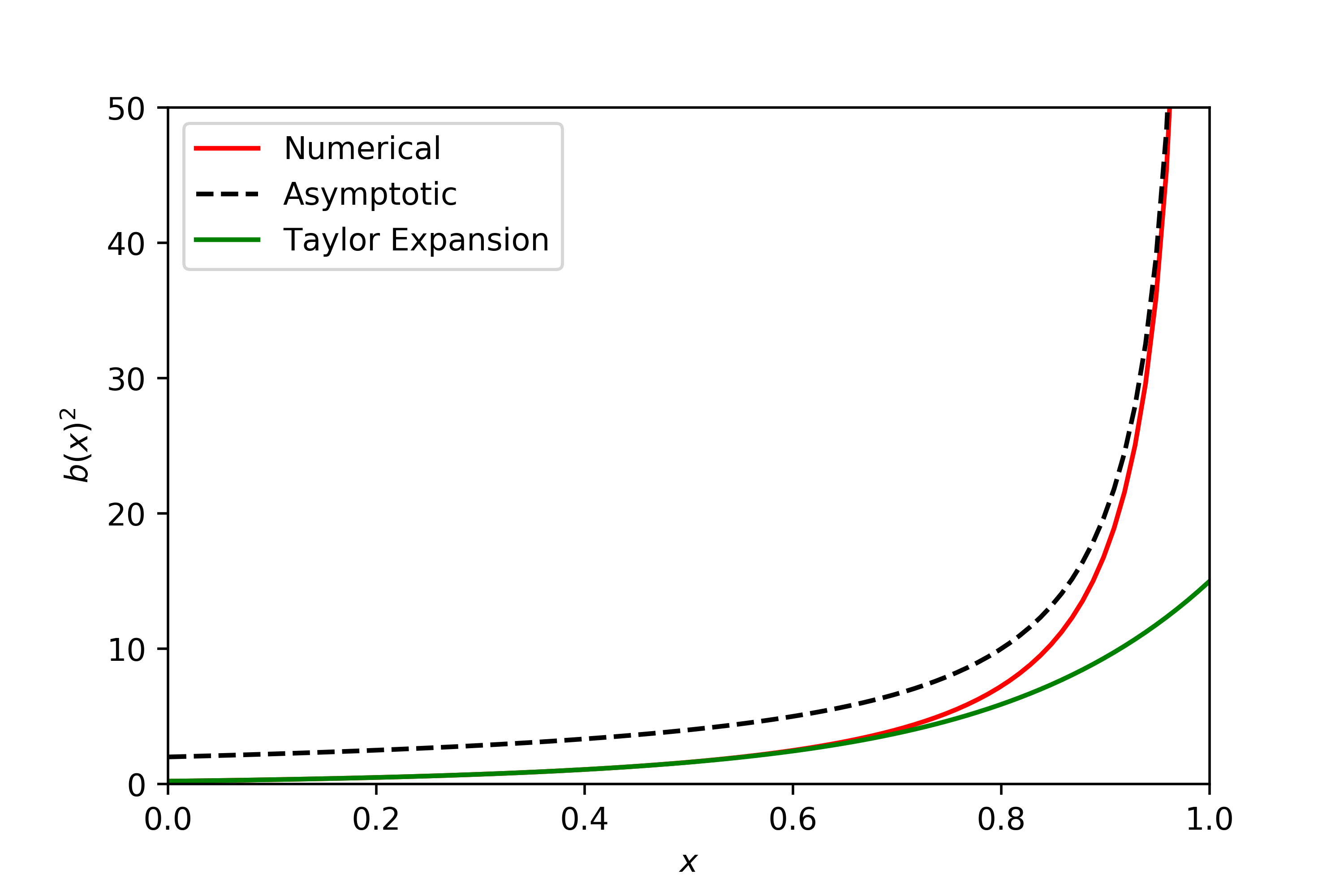}
            \caption[bsq]%
            {{\small $b(x)^2$}}    
            \label{fig:mean and std of net34}
        \end{subfigure}
        \hfill
        \begin{subfigure}[b]{0.475\textwidth}   
            \centering 
            \includegraphics[width=\textwidth]{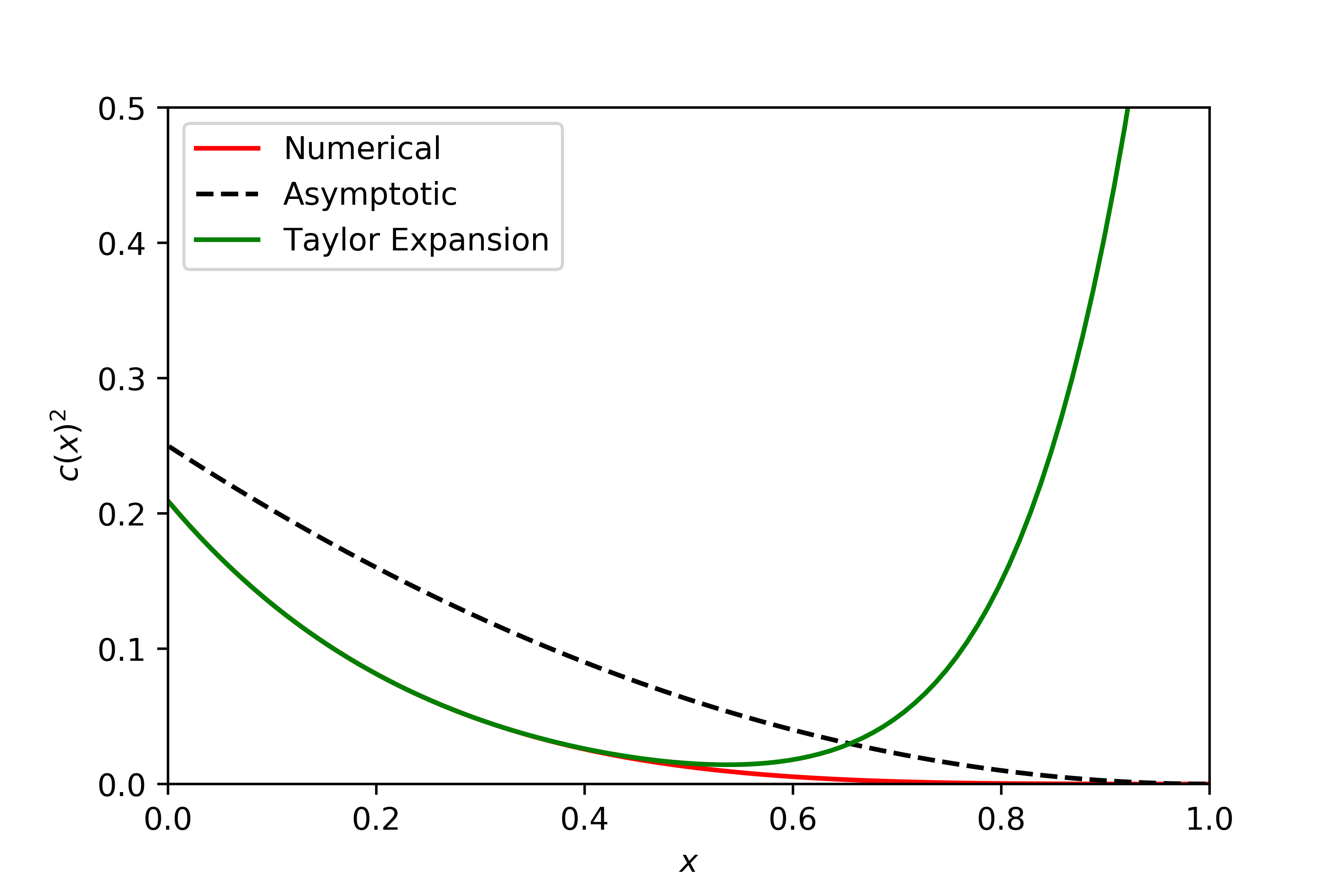}
            \caption[csq]%
            {{\small $c(x)^2$}}    
            \label{fig:mean and std of net44}
        \end{subfigure}
        \caption[ ]
        {\small The four Sutcliffe metric components computed numerically (red) to a precision of $10^{-5}$, the asymptotic expansion given by equation \eqref{eq:sutInf} (dashed) valid in the large x limit and the octic Taylor expansion valid near $x=0$.} 
        \label{fig :: Comparison Sutcliffe}
    \end{figure}

Sutcliffe computed the resulting metric explicitly by computing the abelian gauge fields \eqref{eqn :: boundary gauge field} and integrating over the $S^2$ boundary, however, the resulting expressions are complicated and so we have instead calculated these integrals numerically. In \cite{sutcliffe2022hyperbolic} asymptotic expansions and Taylor expansions were computed. A comparison with our numerical evaluations is presented in figure \ref{fig :: Comparison Sutcliffe}.

In the coordinates of equation \eqref{eqn :: the spectral curves}, near the bolt, $r=0$, we have 
\begin{align}
    g_{S} \sim A(dr^2 + 4r^2\sigma_1^2) + B(\sigma_2^2 + \sigma_3^2)
\end{align}
for some constant $A,B$, where $\sigma_i$ are the left-invariant one-forms of $SU(2)$. Near $r = 1$, 
\begin{align}\label{eq:sutInf}
    g_{S} \sim \frac{1}{2(1-r)^2}dr^2 + \frac{2}{1-r}(\sigma_1^2 + \sigma_2^2) + \frac{(1-r)^2}{4}\sigma_3^2
\end{align}
In this limit, the metric approaches a metric on a circle bundle over $(0,1)\times S^2$ with the $S^1$ metric component vanishing exponentially. This is in contrast to the Atiyah-Hitchin metric where the $S^1$ factor doesn't decay.

\subsection{Hitchin's Metric}
In \cite{tod1994self} Tod considered the conformal class of four dimensional Bianchi IX metrics with vanishing self-dual Weyl curvature and $SO(3)$ symmetry. He showed that these can be described by a variable $x$ and functions $\Omega_1(x),\Omega_2(x),\Omega_3(x)$

\begin{align}
    g_0 = \frac{dx^2}{x(1-x)} + \frac{\sigma_1^2}{\Omega_1^2} + \frac{\sigma_2^2(1-x)}{\Omega_2^2} + \frac{\sigma_3^2x}{\Omega_3^2}
\end{align}
satisfying
\begin{align}
    \Omega_1' &= -\frac{\Omega_2\Omega_3}{x(1-x)}\\
    \Omega_2' &= -\frac{\Omega_1\Omega_3}{x}\\
    \Omega_3' &= -\frac{\Omega_2\Omega_1}{(1-x)}
\end{align}
and
\begin{align}\label{constraintOmega}
    \Omega_1^2 - \Omega_2^2 - \Omega_3^2 = -\frac{1}{4}.
\end{align}
Crucially this system of equations can be shown to be equivalent to the Painleve VI equations.

Tod also showed that there is a conformal rescaling $g = \Theta g_0$ which is Einstein, where 
\begin{align}
\Theta &= ND^{-2}\\
N&=2\Omega_1\Omega_2\Omega_3(4x\Omega_1\Omega_2\Omega_3 + P)\\
P &= x(\Omega_1^2 + \Omega_2^2) - (1-4\Omega_3^2)(\Omega_2^2 - (1-x)\Omega_1^2)\\
D &= x\Omega_1\Omega_2 + 2\Omega_3(\Omega_2^2 - (1-x)\Omega_1^2).
\end{align}
Exploiting the twistor theory of self-dual Einstein 4-manifolds, Hitchin was able to solve this system of ODEs exactly, resulting in the following family of metrics for each integer $k\ge 3$ and with $x\in [1,\infty)$. The metric has the asymptotic form 

\begin{align}\label{nearzero}
    g_{H} \sim A\left(\frac{dx^2}{x^{4}} + 4(1-1/x)^2\sigma_2^2 \right) + B(\sigma_1^2 + \sigma_3^2).
\end{align}
near $x = 1$, or if we change coordinates to $\tau=1-1/x$

\begin{align}
    g_{H} \sim A\left(d\tau^2 + 4\tau^2\sigma_2^2 \right) + B(\sigma_1^2 + \sigma_3^2).
\end{align}
for some constants $A,B$. Therefore the metric extends continuously over the $x=1$ bolt. Conversely, as $x\to\infty$

\begin{align}\label{nearinfty}
    g_{H} \sim \frac{dx^2}{x^{2+2/k}} + \frac{4k^2\sigma_3^2}{(k-2)^2x^{2/k}} + D(\sigma_1^2 + \sigma_2^2).
\end{align}
and after a change of coordinates $t= x^{-1/k}$ becomes

\begin{align}\label{hitRad}
    g_H = dt^2 + \frac{4k^2t^2}{(k-2)^2}\sigma_3^2 + D(\sigma_1^2 + \sigma_2^2)
\end{align}
for some constant $D$. With the exception of $k=4$, these metrics have a conical singularity at $t = 0$.

In solving this system, Hitchin showed that this metric, provided $k\ge 5$, is naturally a metric on the moduli space of centred hyperbolic 2-monopoles with mass $m = (k-4)/4$. We briefly describe this connection but refer to \cite{hitchin1993new} for details. Associated with a self-dual Einstein four-manifold $M$ is a complex three-dimensional twistor space $Z$. The points of $M$ correspond to projective lines in $Z$ with normal bundle $\mathcal{O}(1)\oplus \mathcal{O}(1)$ and $Z$ has a real structure with no real points. Let $\pi : \C P^1\times\C P^1 \rightarrow \C P^2$ be the usual 2-1 branched cover, Hitchin constructs $Z$ as the total space of the projective bundle $p : P(V_k) \rightarrow \C P^2$, where $V_k = \pi_*\mathcal{O}_{\C P^1\times\C P^1}(k,0)$. These projective lines lift to elliptic curves in the $\C P^1 \times \C P^1$ cover; it is these curves, Hitchin shows, that are the spectral curves of hyperbolic monopoles.

\section{Conformal (in-)equivalence}\label{sec :: conf ineq}
In this section we make a direct comparison of the two metrics in the special case where $k=6$ in Hitchin's metric, then each metric corresponds to monopoles of $m=1/2$ and sectional curvature $-1$. In what follows we show that they are not conformally equivalent. In order to do this we leverage the fact that they can each be understood as metrics on the moduli space of 2-monopole spectral curves. Then, by comparing their underlying spectral curves, we find a common coordinate system.

Sutcliffe gives his spectral curves explicitly; parametrized by a real parameter $r\in [0,1)$, they take the form of equation \eqref{eqn :: the spectral curves} and $SO(3)$ orbits.

\begin{figure}[ht]
\centering
    \includegraphics[width=.45\textwidth]{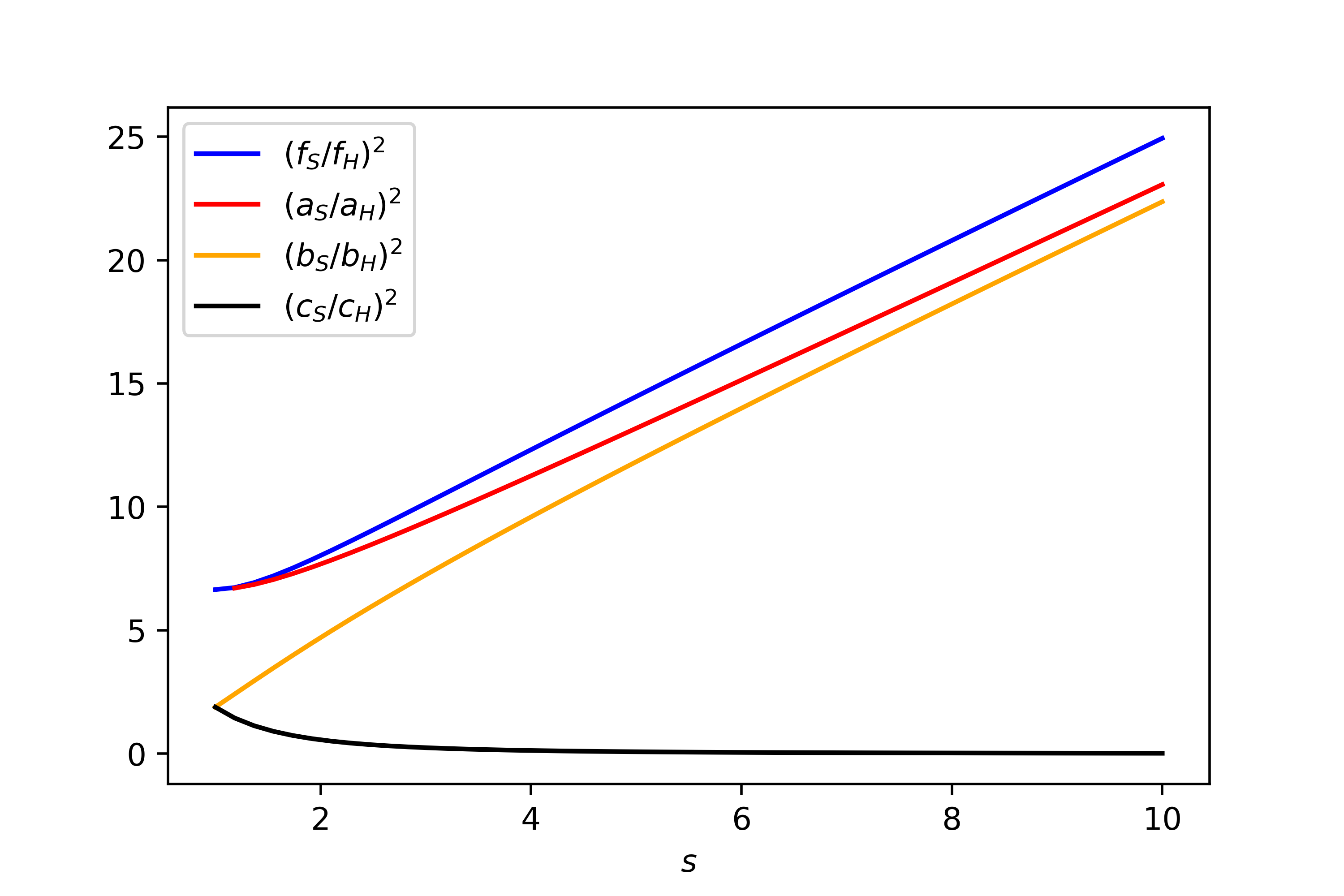}\hfill
\caption{A plot of the ratio of the respective metric components.}\label{fig:MetricComparison}
\end{figure}

Hitchin does not explicitly give the spectral curves corresponding to his metric. However, for the $k=6$ case the monopole mass is $m=1/2$ and so the spectral curves can be identified with circle invariant instantons with complex ADHM data. Hitchin gives this data (page 220 of \cite{hitchin1993new}). The spectral curve can be computed from the complex ADHM data \cite{murray2000complete,bolognesi2014hyperbolic} 
\begin{align}
    \frac{-s+1}{s+1}(\eta^2\zeta^2 + 1) + (\eta^2 + \zeta^2) - \frac{4s}{(1+s)^2}\eta\zeta = 0
\end{align}
with $s\in[1,\infty)$. Therefore, we can parametrize both metrics using Hitchin's parameter $s$ since
\begin{align}
    r = \frac{s-1}{s+1}.
\end{align}
In this common coordinate system each can be described by a single parameter $s\in [1,\infty)$ as metrics of the form
\begin{align}\label{eq:BianchiIX}
    g_X = f_X(s)^2ds^2 + a_X(s)^2\sigma_1^2+b_X(s)^2\sigma_2^2+c_X(s)^2\sigma_3^2
\end{align}
where the subscript $X=H,S$ for Hitchin's and Sutcliffe's metrics respectively. Since we have found a common coordinate parameter $s$, if the two metrics are conformally equivalent the conformal factor can be found from ratios of each corresponding metric components. But as can be seen from the plot in figure \ref{fig:MetricComparison} the metrics cannot be conformally related. 

This agrees with the asymptotic formulae \eqref{eq:sutInf},\eqref{hitRad} as both $a_H/a_S \sim b_H/b_S \sim \mathcal{O}(s)$; however $c_S/c_H \rightarrow 0$, showing there is not even asymptotic conformal equivalence. This strongly suggests that the metrics cannot both be good models of monopole dynamics. We expand on this in the next section.

\begin{figure}[ht]
\centering
    \includegraphics[width=.45\textwidth]{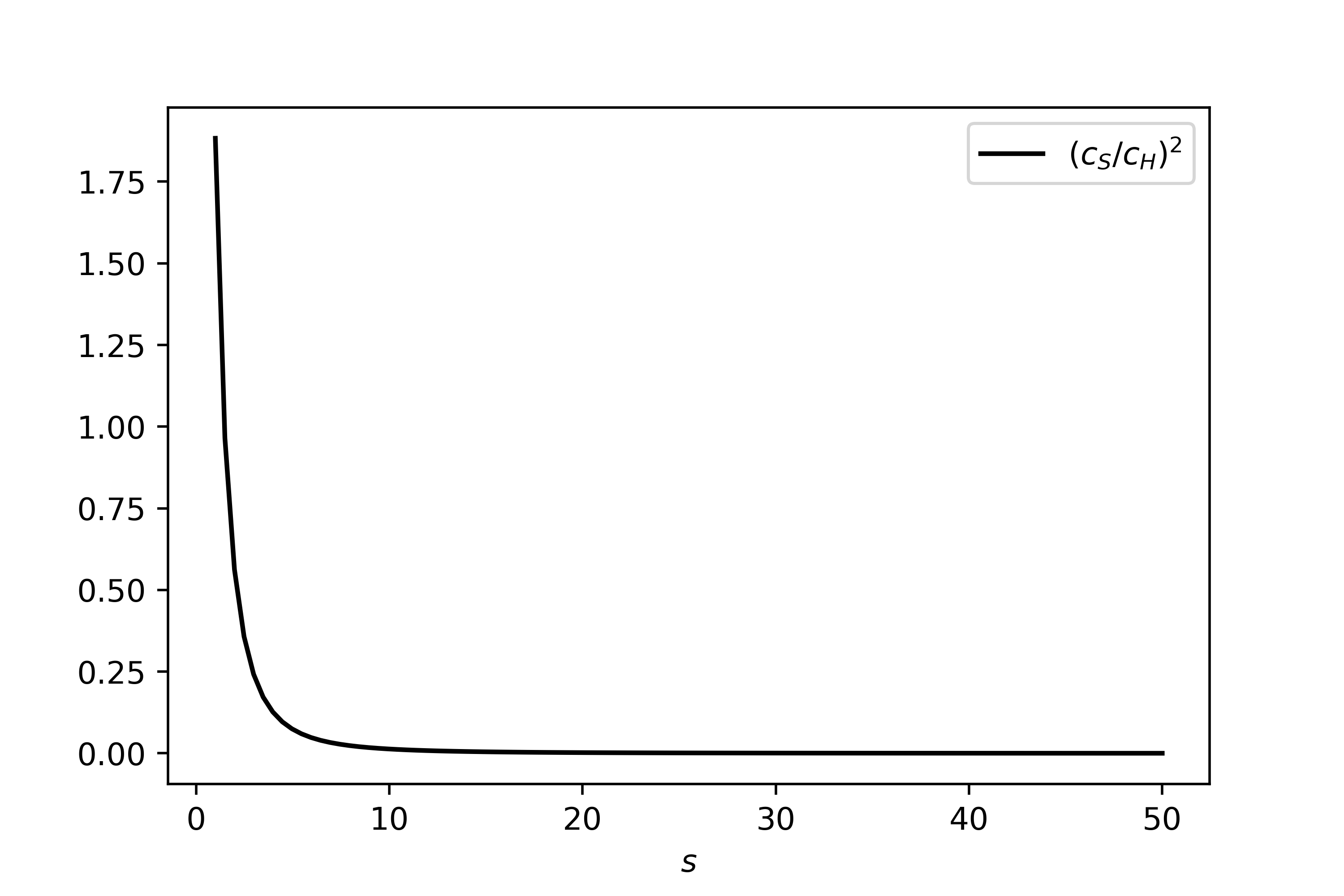}\hfill
\caption{ The ratio of the final components corresponding to the relative phase of the Sutcliffe and Hitchin metrics.}\label{fig:: C}
\end{figure}

\begin{figure}[ht]
\centering
    \includegraphics[width=.45\textwidth]{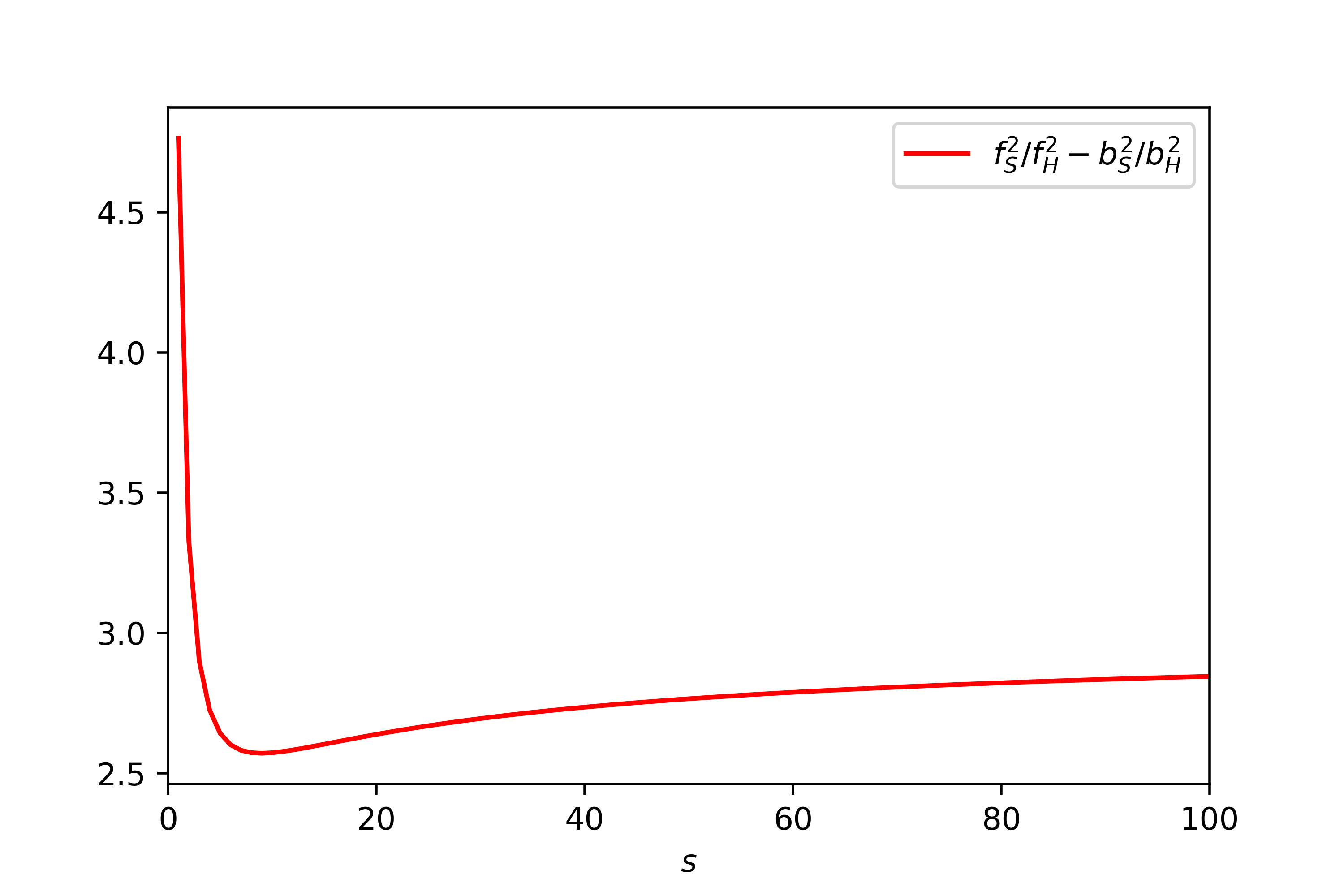}\hfill
\caption{The difference of the ratio of the first metric components and the third metric components  of the Sutcliffe and Hitchin metrics in the form of Bianchi IX \eqref{eq:BianchiIX}.}\label{fig::diff}
\end{figure}

The component in Sutcliffe's $c_S$ rapidly decays to zero whereas the corresponding component in Hitchin's metric approaches a constant. This can be seen in the rapid decay in figure \ref{fig:: C}.
It is interesting to consider whether the other 3 components could be asymptotically conformal as these components correspond to the motion of monopole positions in the underlying hyperbolic space. Unfortunately, $(f_H/f_S)^2 - (b_H/b_S)^2$ does not converge to zero; as can be seen in figure \ref{fig::diff}, the remaining 3 metric components are not asymptotically conformal either.
 
\section{Point Particle approximation}\label{sec:: pointp}
Hyperbolic monopole dynamics are thought to be
approximated in the large separation limit by a metric found from point particle dynamics. Motivated by this, Franchetti and Ross derived such a metric
\cite{franchetti2023asymptotic}. For 2-monopoles they found 
\\
\begin{align}
    g_{hTN} &= Vg_{H^3} + 4M^2V^{-1}\sigma_3^2\\
    g_{hTN} &= V\left(dr^2 + L^2\sinh^2\left(\frac{r}{L}\right)(\sigma_1^2+\sigma_2^2)\right) + 4M^2V^{-1}\sigma_3^2 \quad\text{where}\\
    V &= 1 + \frac{4M}{L}\left(e^{\frac{2r}{L}}-1\right)^{-1}\label{eq:hTNPot}
\end{align}
\\
where $r$ is the radial geodesic coordinate of $H^3$, $M$ is known as a mass parameter and $L$ the radius of curvature of hyperbolic space. This is also known as the Hyperbolic Taub-NUT metric (hTN). Which again can be written in Bianchi IX form as in equation \eqref{eq:BianchiIX}.

We therefore wish to compare three metrics in some large separation limit. For Hitchin's metric we need the asymptotic form given by equation \eqref{nearinfty} and for Sutcliffe's metric we take \eqref{eq:sutInf}. We now find a common coordinate system and conformal scale.

First, conformally rescale so that the coefficient of $\sigma^2_3$ is $1$, then re-parametrize the radial coordinate so that the $\sigma_1,\sigma_2$ components are the same and asymptotically linear -- we name the new variable '$z$'. Label the new metrics $g'_{hTN},g'_{H},g'_{S}$ accordingly. Then,
\begin{align}
    g'_H &\sim c_1z^{-2}dz^2 + z(\sigma_1^2+\sigma_2^2) + \sigma_3^2\\
    g'_S &\sim c_2z^{-{\frac{4}{3}}}dz^2 + z(\sigma_1^2 + \sigma_2^2) + \sigma_3^2\label{SutRes}\\
    g'_{hTN} &\sim c_3z^{-2}dz^2 + z(\sigma_1^2+\sigma_2^2) + \sigma_3^2
\end{align}
as demonstrated by Franchetti and Ross, the constants $c_1$ and $c_3$ can be made to agree by a correct choice of the parameters $L,M$ in equation \eqref{eq:hTNPot}. Sutcliffe's metric fails to agree with the other two in this limit as can be seen by the $-4/3$ exponent in \eqref{SutRes}. Therefore only Hitchin's metric agrees asymptotically and up to a conformal factor with the point particle approximation.

\section{Dynamics}\label{sec:: dyn}
Sutcliffe's metric has the advantage of being more directly interpretable in terms of monopole dynamics. Although Sutcliffe's metric is not asymptotically conformal to the hyperbolic Taub-NUT metric, we will show that it does possess some similarities when one considers monopole locations only. We begin by relating the spectral curve parameter $r$ to the geodesic distance $\rho$ of the Higgs field zeros to the origin of the base space, as at large separation the Higgs field zeros can be interpreted as individual monopole positions. The relationship between $r$ and $\rho$ was found exactly by Sutcliffe and is given by equation (3.3) in \cite{sutcliffe2022hyperbolic}. We reproduce this in an approximate form by considering the asymptotic metric equation \eqref{eq:sutInf}. This metric should be asymptotic to a circle bundle over the monopole base space $H^3$.

The asymptotically hyperbolic metric in geodesic normal coordinates is given by
\begin{align}
   g_{H^3}\sim d\rho^2 + \frac{1}{4}e^{2\rho}(\sigma_1^2+\sigma_2^2)
\end{align}
to relate this to equation \eqref{eq:sutInf} we make the substitution $2\rho \sim -\ln(\frac{1}{4}(1-r))$ giving
\begin{align}\label{eqn :: sutcliffe geodesic asymptote}
     g_S &\sim 2\left(d\rho^2 + \frac{1}{4}e^{2\rho}(\sigma_1^2+\sigma_2^2)\right) + 4e^{-4\rho}\sigma_3^2\\
     &\sim2g_{H^3} + 4e^{-4\rho}\sigma_3^2.
\end{align}
By introducing Euler angles we write this metric in a more familiar form
\begin{align}\label{eq:: asymptotic sutcliffe metric magnetic-geodesic}
    g_S \sim 2g_{H^3} + 4e^{-4\rho}(d\varphi - A)^2
\end{align}
where $A = - \cos{\theta}d\psi$ and $g_{H^{3}}=(\frac{2}{1-|x|^2})^2d\mathbf{x}\cdot d\mathbf{x}$. This corresponds to point particles at position $\pm x$ with relative phases $\varphi$. We are interested in geodesics, so we consider the energy functional

\begin{align}
    E = \int 2 g_{ij}\dot{x}^i\dot{x}^j + 4e^{-4\rho}(\dot{\varphi} - A_i\dot{x}^i)^2
\end{align}
where $g_{ij} = (g_{H^3})_{ij}$. This is invariant under translations of $\varphi$ leading to the conserved charge
\begin{align}\label{eq :: SutcliffeCharge}
    Q = 2e^{-4\rho}(\dot{\varphi} - A\cdot \dot{\mathbf{x}}).
\end{align}
We can consider variations in $\mathbf{x}$ of the energy functional giving Euler-Lagrange equation
\begin{align}\label{eqn :: magnetic geodesic}
    \frac{d}{dt}\left(g_{ij}\dot{x}^j\right) - \frac{1}{2}\pd{g_{kj}}{x_i}\dot{x}^k\dot{x}^j + QF_{ij}\dot{x}^j = 0
\end{align}
where $F = dA$. This is the so called magnetic-geodesic equation, corresponding to the geodesics of \eqref{eq:: asymptotic sutcliffe metric magnetic-geodesic}. We now compare this with the point particle approximation of Franchetti and Ross by performing the same calculation on hyperbolic Taub-NUT metric. In this case we have
\begin{align}\label{eqn:: asym htn}
    g_{hTN} \sim g_{H^3} + (d\varphi - A)^2
\end{align}
by considering the geodesic energy functional, the conserved quantity coming from $\varphi$ invariance is
\begin{align}
    Q_{hTN} = (\dot{\varphi} - A\cdot \dot{\mathbf{x}}).
\end{align}
 The magnetic geodesic equation for hTN takes the same form as \eqref{eqn :: magnetic geodesic} but with $Q$ replaced by $Q_{hTN}$. If we equate the parameters $Q,Q_{hTN}$ then the resulting magnetic-geodesics in $H^3$, i.e. solutions to \eqref{eqn :: magnetic geodesic}, will be the same. In this case the geodesics will be identical in the spatial dimensions and the remaining degree of freedom corresponds to the relative phase of the monopoles. As can be seen from \eqref{eq :: SutcliffeCharge}, $\dot{\varphi}$ will grow exponentially in geodesic radial distance. This is in contrast to the hTN metric whose relative phase will grow linearly.

There is a natural family of solutions to \eqref{eqn :: magnetic geodesic} which correspond to the radial geodesics of the base hyperbolic 3-space, where radial here means the radial lines through the origin of the ball model. To see this note that the quantity $F_{ij}\dot{x}_j$ vanishes along radial geodesics as $F = dA = \sin{\theta} d\theta \wedge d\psi$. Then the remaining equation is just the geodesic equation for the base hyperbolic space
\begin{align}
    \frac{d}{dt}\left(2g_{ii}\dot{x}^i\right) - \pd{g_{kj}}{x_i}\dot{x}^k\dot{x}^j = 0.
\end{align}
Then the solutions in terms of the geodesic radial distance for the hyperbolic Taub-NUT metric take the form
\begin{align}
    (\tanh{(\rho/2)} u_1,\tanh{(\rho/2)} u_2,\tanh{(\rho/2)} u_3,\rho Q)
\end{align}
for some fixed $(u_1,u_2,u_3)\in S^2 = \partial H^3$. For Sutcliffe's metric we find, 
\begin{align}
    (\tanh{(\rho/2)} u_1,\tanh{(\rho/2)} u_2,\tanh{(\rho/2)} u_3,\frac{1}{8}Qe^{\rho})
\end{align}
This clearly demonstrates the disparity in the relative phase of Sutcliffe's metric and the hTN metric.

\section{Harmonic Forms}\label{sec:: harm}
Other objects of interest in relation to monopole dynamics are the $L^2$-harmonic 2-forms. These are the bound states of monopoles after quantization. An important conjecture in this direction is the Sen conjecture, proposed by Ashoke Sen in \cite{sen1994dyon}. We state the charge two case:
\begin{conjecture}\label{Sen}
    Let M denote the moduli space of (centred) magnetic 2-monopoles, then the spaces of $L^2$ harmonic $p$-forms obey
    \begin{align}
        \mathcal{H}_{(2)}^p = \begin{cases} 
      \mathbb{C} & p = 2 \\
      0 & \text{otherwise.} 
   \end{cases}
    \end{align}
\end{conjecture}
This has been confirmed for the Atiyah-Hitchin metric of Euclidean 2-monopoles by Hitchin in \cite{hitchin1993new}.

\subsection{$L^2$ Cohomology}

To motivate the Hodge theory of non-compact manifolds we first consider the much more straightforward theory on compact manifolds. Let $M$ be a smooth, compact Riemannian manifold without boundary. Then we have the Hodge decomposition of smooth $k$-forms
\begin{align}
    \Omega^k(M) = \text{im}{d_{k-1}}\oplus \text{im}{\delta_{k+1}}\oplus \mathcal{H}^k(M)
\end{align}
where the map $\delta_{k+1} : \Omega^{k+1}(M) \to \Omega^k(M)$ is the usual co-differential operator and $\mathcal{H}^k$ the kernel of the Laplace-Beltrami operator on $M$. From the Hodge decomposition we have
\begin{align}
    \ker{d_{k+1}} = \text{im}{d_{k-1}}\oplus \mathcal{H}^k(M).
\end{align}
The orthogonality of these spaces follows from Stokes' theorem and so it follows that
\begin{align}
    H^k(M) = \frac{\ker{d_{k}}}{\text{im}{d_{k-1}}} = \mathcal{H}^k(M)
\end{align}
giving the familiar correspondence between de Rham cohomology and harmonic forms. We now relax the assumption of compactness and restrict our attention to $L^2$ forms. This restriction allows us to make a connection to Sen's conjecture via a Hodge theorem for $L^2$ forms. $L^2$ forms enjoy a similar Hodge type decomposition due to Kodaira \cite{kodairaharmonic1949}
\begin{align}
    L^2 = \overline{d\Lambda_c^{p-1}}\oplus \overline{\delta\Lambda_c^{p+1}}\oplus \mathcal{H}_{(2)}^k(M)
\end{align}
where $\Lambda_c^{k}$ are smooth forms of compact support and $\mathcal{H}_{(2)}^k(M)$ denotes the space of $L^2$ forms which are closed with respect to both $d$ and $\delta$. We will need to restrict the domain operators $d$ and $\delta$ to $\text{dom}(d) :=\{a \in L^2\mid da \in L^2\}$ and similarly for $\text{dom}(\delta)$. Next, we briefly define $L^2$ cohomology, for details see the introductory article by Dai \cite{dai2011introduction}. We can then define the $L^2$ cohomology as
\begin{align}
    H^k_{(2)}(M) = {\ker{d_{k}}}/{\text{im}{d_{k-1}}}.
\end{align}

There is another related space which comes from considering the strong closure of the operator $d$, denoted $\bar{d}$. Denote by $\text{dom}(d)$ the forms $\alpha \in L^2$ such that $da\in L^2$. Given an $L^2$ form $\alpha$ if there exists a sequence of $L^2$ forms $\eta_i \in \text{dom}{(d)} \rightarrow \alpha$ then $\alpha$ is said to be in the domain of $\bar{d}$ if $d\eta_i \rightarrow \xi \in L^2$ and we define $\bar{d}\alpha := \xi$. The set
\begin{align}
    {\ker{\bar{d}_{k}}}/{\text{im}{\bar{d}_{k-1}}}
\end{align}
is known to be isomorphic to $H^k_{(2)}(M)$\cite{cheeger1980hodge} so we take this as our definition of the $L^2$ cohomology. There is a problem with our definition of $L^2$ cohomology in how it relates the harmonic forms via the Kodaira decomposition defined above. Namely, it is not always true that $\overline{d\Lambda_c^{k-1}} \subset {\text{im}{\bar{d}_{k-1}}}$ and so the induced map $H_{(2)}^k(M) \rightarrow \mathcal{H}_{(2)}^k(M)$ is not necessarily surjective. To circumvent this we often make use of a smaller space, namely the reduced $L^2$ cohomology: 
\begin{align}
\bar{H}^k_{(2)} =  {\ker{\bar{d}_{k}}}/\overline{\text{im}{\bar{d}_{k-1}}}
\end{align}
We are then left with a surjective map $\bar{H}_{(2)}^k(M) \rightarrow \mathcal{H}_{(2)}^k(M)$. Injectivity can be deduced easily from an $L^2$ Stokes' theorem, as one can then show that the spaces $\overline{\text{im}{\bar{d}_{k-1}}}$ and $\mathcal{H}_{(2)}^k$ are orthogonal. Therefore establishing an $L^2$ Stokes' theorem for Hitchin's manifold is the first priority.
\subsection{ The $L^2$ Cohomology of Hitchin's Metric}
%Complete Riemannian manifolds enjoy an $L^2$ Hodge theorem in the sense that the space of $L^2$ harmonic forms are naturally isomorphic to the reduced $L^2$ cohomology: $\bar{H}^p_{(2)}$. Let $\Omega_{(2)}^p(M)$ be the smooth $L^2$ p-forms. The groups $\bar{H}^p_{(2)}$ are defined as the kernel of $d$ modulo the closure of $\text{Im}(d)\cap \Omega_{(2)}^p(M)$. Often the $L^2$ cohomology classes give insight into the structure of the Harmonic forms, this is exemplified by a theorem of Hitchin stated below \ref{thm :: Hit3}.

%Unfortunately, the metric $g_H$ is not complete and so it is not necessarily true that $\mathcal{H}_{(2)}^p \cong \bar{H}_{(2)}^p$. It is well known that for a smooth (possibly incomplete) Riemannian manifold it is sufficient that the $L^2$ Stokes' theorem is true. Therefore we seek to prove

We now specialize to the case of Hitchin's manifold $(M,g_H)$ which is incomplete due to the presence of the conical singularity. Due to the non-trivial fibre structure of Hitchin's manifold, the type of singularity is known as a simple edge singularity, which we define below. The goal of this section is to prove that Hitchin's metric has a Hodge theorem with respect to its (reduced-)$L^2$ cohomology, this is summarized in proposition \ref{prop :: Hitchin Hodge}. As explained earlier, we first need to prove this manifold enjoys a Stokes' theorem on the space $\Omega_{(2)}^\bullet(M)$ of $L^2$ forms. We achieve this by appealing to a theorem of Hunsicker and Mazzeo found in \cite{hunsicker2005harmonic} though we more closely follow the conventions of \cite{behrens2009l2}.
\begin{proposition}\label{prop :: Hitchin Hodge} Let $(M,g_H)$ be Hitchin's manifold, then there exists an isomorphism
\begin{align}
    \bar{H}^p_{(2)}\cong \mathcal{H}^p_{(2)}.
\end{align}
\end{proposition}
Which by the discussion of the previous section is equivalent to proving the following proposition:    
\begin{proposition}\label{prop:hitStokes}
     The incomplete manifold $(M,g_H)$ is such that for all $\alpha \in \Omega^{p-1}_{(2)}$ such that $d\alpha \in \Omega^{p}_{(2)}$ and for all $\beta \in \Omega^{p+1}_{(2)}$ such that $\delta\beta\in \Omega^{p}_{(2)}$
     \begin{align}
         \langle d\alpha,\beta \rangle = \langle \alpha,\delta\beta \rangle.
     \end{align}
\end{proposition}

To define a simple edge singularity we must first define the notion of quasi-isometry

\begin{definition}
       Let $M$ be a smooth manifold. A pair of metrics $g$ and $h$ on $M$ are said to be quasi-isometric if there exists a constant $C\ge 1$ such that
    \begin{align}
        C^{-1}g(X,X) \le h(X,X) \le Cg(X,X).
    \end{align}
    
\end{definition}

\begin{definition}
    A Riemannian manifold $(M,g)$ is said to have a simple edge singularity modelled on the fibre bundle $Y \xrightarrow[]{\pi} B$ with fibre $F$ if there exists an open subset $U \subset M$ such that $M\backslash U$ is a smooth compact manifold and $U$ is quasi-isometric to the cylinder $I\times Y$ with metric
    \begin{align}
        dx^2 + \pi^*g^B + x^2\kappa
    \end{align}
    where $\kappa$ restricts to a metric on each fibre $F$.
\end{definition}
This describes Hitchin metric, which is quasi-isometric to \eqref{hitRad} in an open neighbourhood $\rho = 0$ where it takes the form of fibre bundle over $S^2$ with fibre given by an cone $\mathcal{C}S^1$.

We now state the theorem of Hunsicker and Mazzeo \cite{hunsicker2005harmonic}.
\begin{theorem}
    Let $(M,g)$ be a manifold with a simple edge singularity modelled on the fibre bundle $Y \rightarrow B$ with fibre $F$.
    \begin{enumerate}
        \item If $\dim{F} = 2k-1$ then the $L^2$ Stokes' theorem holds on M.
        \item If $\dim{F} = 2k$ and $H^k(F,\C) =0$, then the $L^2$ Stokes' theorem holds on M.
    \end{enumerate}
\end{theorem}
As described above Hitchin's metric is a simple edge singularity modelled on a fibre bundle with fibre $F=S^1$, therefore by the above theorem Hitchin's metric has a $L^2$ Stokes' theorem. This proves proposition \ref{prop :: Hitchin Hodge} and so the reduced $L^2$ cohomology is isomorphic to the space of $L^2$ harmonic forms for Hitchin's metric.
The following theorem due to Hitchin \cite{hitchin20002} allows us to further restrict the possible $L^2$ harmonic forms.

\begin{theorem}[Hitchin]\label{thm :: Hit3}Let M be a complete Riemannian manifold and let $G$ be a connected Lie group of isometries such that the associated Killing vector fields satisfy
    \begin{align}\label{eq :: linear growth}
        \norm{X(x)} \le c'\rho(x_0,x) + c''
    \end{align}
    where $\rho$ is the geodesic distance. Then each $L^2$ cohomology class is fixed by $G$.
\end{theorem}
The condition \eqref{eq :: linear growth} is commonly referred to as linear growth. The proof of the above theorem does not make use of the completeness condition so we may apply it freely to Riemannian manifolds with a conical singularity and so can also be applied to Hitchin's metric. We now show how this applies to Bianchi IX metrics. First, consider a generic Bianchi IX metric in radial geodesic coordinates given by
\begin{align}
    d\rho^2 + a(\rho)^2\sigma_1^2 + b(\rho)^2\sigma_2^2 + c(\rho)^2\sigma_3^2
\end{align}
The $SO(3)$ Killing vector fields are given by the right invariant vector fields $R_i$, say, then $\norm{R_i}^2$ will be bounded by some linear combination of the metric coefficients. Therefore, if the metric coefficients are at most quadratic in the geodesic radial distance then the metric satisfies the hypothesis of theorem \ref{thm :: Hit3}. This is certainly true for Hitchin's metric as can be seen in equation \eqref{hitRad}.

However, Sutcliffe's metric fails to meet the hypothesis of theorem \ref{thm :: Hit3}, as in radial geodesic coordinates the metric coefficients grow exponentially in radial geodesic distance as can be seen in equation \eqref{eqn :: sutcliffe geodesic asymptote}. So we do not have the benefit of the Hodge theorem and Hitchin's theorem. In spite of this we compute a symmetric $L^2$ harmonic 2-form in the following section.

\subsection{Computing the Forms}
We now apply the results of the previous section to compute the unique harmonic 2-form for Hitchin's metric and the only $SO(3)$ symmetric harmonic form for Sutcliffe's metric. These forms can be written exactly in terms of the metric components. To find the harmonic two-forms we start by considering metrics taking the general form
\begin{align}
    g = f(r)^2dr^2 + a(r)^2\sigma_1^2+b(r)^2\sigma_2^2+c(r)^2\sigma_3^2
\end{align}
 and we reduce our search to the $SO(3)$ invariant forms:
\begin{align}
    \theta = \sum_{i}(\alpha_{i}(r) dx\wedge\sigma_i + \beta_i(r) d\sigma_i)
\end{align}
we then impose closure and co-closure. Let $\lambda_i^\pm = q_i(r)dr\wedge\sigma_i \pm d\sigma_i$, where $q_i(r)$ is defined $\star d\sigma_i = q(r)dr\wedge\sigma_i$, be the dual and anti-self dual forms respectively. Then,
\begin{align}
\theta &= \sum \theta_i^\pm\quad\text{with}\\
\theta_i^\pm&= d(\beta_i^\pm\sigma_i^\pm)
\end{align}
where $(\beta_i^\pm)' = \pm q_i(r)\beta_i^\pm$ and each $\theta_i^\pm$ are themselves harmonic. Therefore, all invariant harmonic forms can be written as a linear combination of these six. 

We may write these forms more explicitly in terms of the metric coefficients. For example when $i=1$ we have
\begin{align}
    \theta_1^\pm = d\left(\exp\left(\pm\int\frac{fa}{bc}\right)\sigma_1\right)
\end{align}
Moreover, these forms are mutually orthogonal and so we need only check which of these 6 forms are $L^2$. 
\\
\\
Both $g_H,g_{S}$ have a bolt-like coordinate singularity with
\begin{align}
    &f=\sqrt{C} &a=2\sqrt{C}r, &&b=\sqrt{D}, &&c=\sqrt{D}
\end{align}
as $r\rightarrow 0$ for some constants $C$ and $D$, in this limit the only integrable forms are
\begin{align}
    \theta_1^\pm \sim \pm\frac{C}{D}2r\exp\left(\pm \frac{C}{D}r^2\right)dr\wedge\sigma_1 \pm \exp\left(\pm \frac{C}{D}r^2\right)\sigma_2\wedge\sigma_3 
\end{align}
Next we consider Sutcliffe's metric at the other end, where geodesic distance is infinite. In the coordinates of \eqref{eqn :: sutcliffe geodesic asymptote}
\begin{align}
    \theta_1^\pm &\sim \mathcal{O}(\pm\exp{\frac{1}{4}e^{2\rho}})
\end{align}
as $\rho\rightarrow \infty$. So the only $SO(3)$ symmetric $L^2$ and harmonic form is $\theta_1^-$. 
\\
\\
For Hitchin's metric we have to contend with the conical singularity of equation \eqref{hitRad}, in this case the remaining harmonic forms are 
\begin{align}\label{eq:: hithform}
    \theta^\pm_1\wedge\star\theta^\pm_1 = (\pm\alpha t^{\pm\frac{k-2}{2}} + \beta t^{\pm\frac{k-2}{2} - 1})dt\wedge\sigma_1\wedge\sigma_2\wedge\sigma_3
\end{align}
for some constants $\alpha,\beta$ as $t\to 0$. Of these, the only $L^2$ 2-form is the self-dual form $\theta_1^+$ so long as $k\ge 4$. In other words, there are no $L^2$ harmonic 2-forms when $k=3$.

\section{Conclusion}
Currently, there is no hyperbolic analogue of Stuart's theorem and therefore we cannot say with certainty whether or not any metric on the moduli space of hyperbolic monopoles is a good model of hyperbolic monopoles dynamics. However some insight can be drawn by making use of the analogy with Euclidean monopoles. From our analysis of the two metrics, we have shown that neither metric enjoys all of the features we would expect hyperbolic monopoles to possess given what we know from the Euclidean theory. However, each metric carries some merit in this regard. 

Sutcliffe's metric has a more natural interpretation in terms of dynamics of gauge fields, though the fields in question are those at infinity rather than the monopole fields in the bulk. However, Sutcliffe's metric fails to be consistent with the point particle approximation given by the hTN metric, unless one ignores the relative phase of the monopoles. Assuming hTN is the correct definition of hyperbolic point particle dynamics, this would rule out Sutcliffe's metric as an accurate description of hyperbolic monopole dynamics. 

An issue with Hitchin's metric is that it does not have an obvious interpretation in terms of dynamics of gauge fields. This problem is perhaps remedied by the methods of Franchetti and Harland in \cite{franchetti20242} using their new gauge fixing condition. From the work of Franchetti and Ross, Hitchin's metric can be made to agree, at least after a conformal rescaling, with the point particle picture.

Whilst neither metric appears to have a completely satisfactory description of the classical dynamics of monopoles they do both appear to have some consistency with what is predicted about their quantum mechanics. Assuming of course that hyperbolic monopoles share a similar quantization scheme to that of their Euclidean counterparts. This is particularly salient to Hitchin's metric which could describe dynamics after a conformal rescaling and since two-forms in dimension 4 are conformally invariant it might be that the two-form \eqref{eq:: hithform} is the correct one for hyperbolic monopole quantum mechanics. 

\bibliography{ref}
\bibliographystyle{ieeetr} 
\end{document}